\documentclass{optica-article}
\journal{opticajournal}
\articletype{Research Article}
\usepackage{lineno}

\usepackage[normalem]{ulem}
\usepackage{cancel}

\providecommand{\keywords}[1]{%
  \par\noindent\textbf{Keywords:} #1
}

\begin{document}

\title{Universal computational super-resolution framework for imaging of quantum dots}

\author{Dominik Va\v{s}inka,\authormark{1,$\dag$} Jaewon Lee,\authormark{2,$\dag$} Charlie Stalker,\authormark{2} Victor Mitryakhin,\authormark{3} Ivan Solovev,\authormark{3} Sven Stephan,\authormark{3,4} Sven H\"{o}fling,\authormark{5} Falk Eilenberger,\authormark{6,7,8} Seth Ariel Tongay,\authormark{9} Christian Schneider,\authormark{3} Miroslav Je\v{z}ek,\authormark{1}  and Ana Predojevi\'{c}\authormark{2,*}}

\address{\authormark{1}Department of Optics, Faculty of Science, Palacký University, 17. listopadu 12, 77900 Olomouc, Czechia\\
\authormark{2}Department of Physics, Stockholm University, 10691 Stockholm, Sweden\\
\authormark{3}Institut of Physics, University of Oldenburg, D-26129 Oldenburg, Germany\\
\authormark{4}University of Applied Sciences Emden/Leer, 26723 Emden, Germany\\
\authormark{5}Technische Physik, Physikalisches Institut and W\"urzburg-Dresden Cluster of Excellence ct.qmat, Universit\"at W\"urzburg, Am Hubland, D-97074 W\"urzburg, Germany\\
\authormark{6}Institute of Applied Physics, Abbe Center of Photonics, Friedrich Schiller University Jena, 07743 Jena, Germany\\
\authormark{7}Fraunhofer-Institute for Applied Optics and Precision Engineering IOF, 07743 Jena, Germany\\
\authormark{8}Max-Planck-School of Photonics, 07743 Jena, Germany\\
\authormark{9}Materials Science and Engineering, School for Engineering of Matter, Transport, and Energy, Arizona State University, Tempe, Arizona 85287, United States}
\email{\authormark{*}ana.predojevic@fysik.su.se}
\authormark{$\dag$}\footnotesize{\textit{These authors contributed equally.}}

\begin{abstract*} 
We present a universal deep-learning method that reconstructs super-resolved images of quantum emitters from a single camera frame measurement. Trained on physics-based synthetic data spanning diverse point-spread functions, aberrations, and noise, the network generalizes across experimental conditions without system-specific retraining. We validate the approach on low- and high-density In(Ga)As quantum dots and strain-induced dots in 2D monolayer WSe$_2$, resolving overlapping emitters even under low signal-to-noise and inhomogeneous backgrounds. By eliminating calibration and iterative acquisitions, this single-shot strategy enables rapid, robust computational super-resolution for nanoscale characterization and quantum photonic device fabrication.
\end{abstract*}

\vspace{0.5em}
\keywords{\textit{Super-resolution imaging, Quantum dots, Deep learning, Convolutional neural network, Calibration-free}}


\section{Introduction}

Quantum dots are nanoscale semiconductor structures with unique optical properties, making them essential for applications in optoelectronic devices, quantum technologies, and biomedical imaging~\cite{Couteau2023_1, Couteau2023_2}. However, the diffraction limit of conventional optical imaging systems restricts the resolution of the fabrication and characterization of complex quantum-dot devices, preventing accurate spatial and spectral analysis at the nanometer scale. To mitigate this, super-resolution imaging techniques have emerged as a powerful tool to overcome this limitation, enabling precise localization and characterization of quantum dots beyond the diffraction barrier.

Traditional super-resolution methods, such as single-emitter localization microscopy, require either sufficiently isolated dots (sparse samples) or specific spectral or temporal behavior (stochastic blinking). When quantum dots and other emitters are observed with non-overlapping point-spread functions, the localization can be achieved down to a few dozen nanometers by various fitting procedures~\cite{Schneider2009, Sapienza2015, Copeland2018, Copeland2024}.
For dense samples, the diffraction-limited images of individual emitters overlap and cannot be localized based on a single camera shot. More complex multi-shot approaches must be adopted in such cases. For example, using spectral emission features allows distinguishing individual quantum emitters~\cite{Liu2024, Musavinezhad2024}. However, this wavelength-selective approach is experimentally demanding and limited by small spectral differences (less than a part of a nanometer) typically present in quantum dot samples.
Alternatively, quantum dot blinking and the acquisition of multiple images reduce the effective sample density, allowing the localization at the expense of a very slow imaging process~\cite{Lidke2005, Dertinger2009, Wang2013}. However, both the spectral distinguishability and stochastic blinking are undesirable for practical applications of quantum dots in quantum technology. This makes the fabrication of quantum dot devices that are based on dense samples specifically problematic due to the need to position more than one quantum dot within the same device and ensure the known spatial distances of the individual dots~\cite{Koshnegar2017}. 

Some of the above-listed issues can be partially solved by the current optical super-resolution methods. However, these require a significant prior knowledge of the imaging system and extensive calibration~\cite{Deschout2014, DECODE2021, Copeland2024}. These constraints present major challenges, particularly in environments with high emitter densities, varying optical conditions, and low signal-to-noise ratios. Deep-learning-based approaches have also been used to improve resolution in various emitter systems~\cite{DeepSTORM2018, DeepSTORM3D2020, DECODE2021, McDonald2019, Impertro2023}, as well as in broader super-resolution imaging modalities~\cite{Zhang2018, Ouyang2018, Kim2019, Kim2022, Midtvedt2022, Kudyshev2023, Chen2023, Chen2023_2, Zhang2025_1, Zhang2025_2}. However, most of these methods require retraining for each specific optical setup, which makes them labor-intensive and system-dependent.

Here, we propose a universal approach to quantum dot computational super-resolution. It also works for dense samples and requires only a single camera image, eliminating the need for spectral distinguishability or stochastic blinking. By using a deep learning model that has been trained on a wide variety of numerically simulated imaging conditions, we can eliminate the need for calibration datasets and the requirements to have explicit knowledge of the optical system parameters. This approach enables us to reconstruct the super-resolved images of quantum dots directly from resolution-limited intensity data in a single shot, thereby improving imaging accuracy while significantly reducing experimental and computational complexity. The ability to adapt to diverse imaging conditions without retraining makes this method particularly suitable for real-time applications in nanophotonics, quantum optics, and materials science. We demonstrate the performance of this universal computational super-resolution imaging method using various semiconductor quantum dot samples. Reconstructions are performed without any prior information on the sample and imaging system. These results pave the way for improved nanoscale characterization and advanced quantum devices development, particularly those containing several tightly packed quantum dots in complex spatial configurations~\cite{Koshnegar2017}.


\section{Universal deep learning model}

\subsection*{Data simulation}
Rather than using experimentally acquired images, we used a simulated dataset to train our universal model. Each data pair consisted of a $50\times 50$ pixel low-resolution image and its corresponding upsampled $200 \times 200$ ground truth representation. These data pairs were generated through a structured process designed to simulate diverse imaging scenarios. First, emitters were positioned within the $200 \times 200$ pixel grid with their intensities drawn from a Poisson distribution. The mean intensity was randomly selected within the~$(1, 10^4)$ photon range to cover various signal-to-noise ratios. This initial grid of point-like emitters served as the ground truth image.

To simulate the limits of optical resolution, we convolved the ground truth with a point spread function~(PSF) of an imaging system. The PSF was randomly chosen to be either a Gaussian function or an Airy disc to simulate high and low numerical aperture imaging, respectively~\cite{Stallinga2010}. The trained model is able to generalize to intermediate Gauss–Airy PSFs and optical aberrations that were not encountered during training, as shown in the supplementary information. The PSF was then further adjusted to introduce more variability. Its full width at half maximum was randomly set within the approximate range of $(8, 40)$ pixels, corresponding to $(2, 10)$ pixels in the $50 \times 50$ grid. To account for deformations, we implemented a PSF asymmetry by applying a squeezing transformation along a random axis. These variations made the model more robust by exposing it to a wider range of optical conditions.

After convolution, the image was downsampled to $50 \times 50$ pixels to simulate the lower resolution of real-world scenarios. Finally, shot noise with average intensities ranging from 1 to 100~photons per pixel was added to the background. Each training sample was simulated independently using a unique, randomly generated combination of optical parameters. Collecting a broad range of synthetic training samples enables the network to generalize across setups without the need for recalibration. 

Furthermore, expanding the range of simulated conditions and incorporating additional imperfections, such as optical aberrations, various noise types, and inhomogeneous backgrounds, would facilitate even greater generalization, making the network adaptable to a broader spectrum of real-world applications. Nonetheless, despite never encountering these effects during training, the model exhibits a high reconstruction ability, as demonstrated in the Results. The real-world measured samples unavoidably suffer from such imperfections, and the reconstruction accuracy highlights the generalization ability that even the simple simulation process can impart.

\subsection*{Deep learning architecture}
\begin{figure}
	\includegraphics[width=0.95\columnwidth]{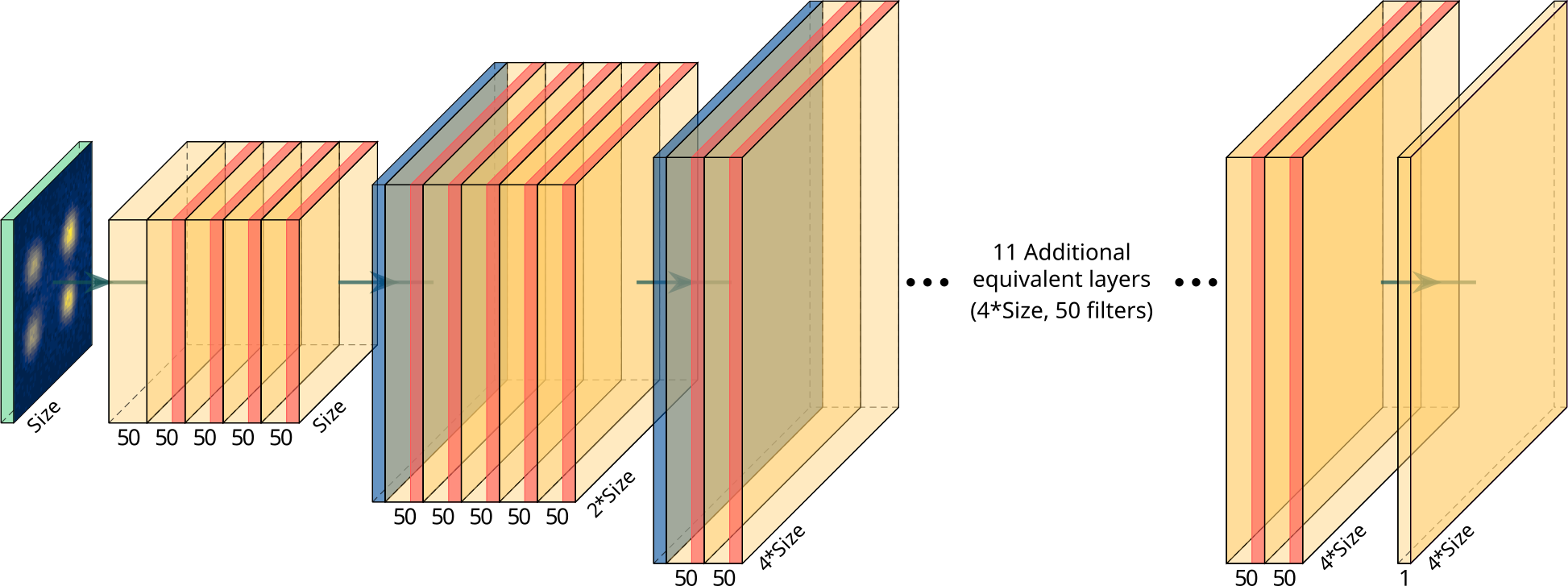}
	\centering
	\caption{
		Visualization of the model architecture. All convolutional layers (yellow) use 50~filters with a Leaky~ReLU activation, except the output layer with a single filter and a softmax activation. Upsampling layers (blue) increase the spatial resolution of feature maps by a factor of two while preserving the number of channels. Dropout (indicated by red stripes) is applied for regularization.
	}
	\label{fig:CNN}
\end{figure}

We have developed a universal deep-learning model for computational super-resolution imaging of quantum dots, designed to operate independently of specific imaging system parameters. This model, illustrated in Fig.~\ref{fig:CNN}, is based on a convolutional neural network that takes a low-resolution intensity image as input and reconstructs a super-resolved output without requiring prior information on the imaging system. Given the convolutional architecture, the network can process images of arbitrary dimensions directly without resizing or cropping. The network features 50 filters for each of its 25 hidden Conv2D layers and extracts spatial structures using a local convolution with trainable $5 \times 5$ kernels. Nonlinearity was introduced via a Leaky~ReLU activation function~\cite{Maas2013} with a negative slope of 0.05, providing efficiency while preventing neuron inactivity from the dying ReLU problem. The final layer utilizes a softmax activation function~\cite{Goodfellow2016} to ensure the non-negativity of the reconstructed intensities and to impose an~L1 normalization constraint on the reconstructed image.

To improve generalization and prevent overfitting, a dropout regularization~\cite{Srivastava2014} with a 0.01~rate was applied to every convolutional layer, except the first and last layers. A crucial component of the model resolution-enhancement capability is the implementation of upsampling layers, which progressively increase the image dimensions through interpolative resizing. Specifically, upsampling layers doubling the image size were placed after the 5th and 10th convolutional layers, resulting in a fourfold magnification at the output (see Fig.~\ref{fig:CNN}). Despite its high performance, the network remains relatively modest, containing approximately 1.5 million trainable parameters, leaving room for improving its robustness and adaptability even further.

The presented model architecture and hyperparameter settings were determined through manual optimization across a wide range of imaging conditions. We explored numerous configurations and observed that performance is strongly influenced by model complexity. While other hyperparameters have a modest effect, the generalization ability improves primarily with increased architecture complexity, i.e., the number of trainable parameters. Further scaling of the model may provide even greater accuracy and generalization, at the cost of higher computational demands during training. Given our available computational resources, we show that even the moderately complex architecture can already achieve strong generability and reconstruction accuracy in real-world applications.

\subsection*{Training}

The model was trained using incremental learning~\cite{Ven2022}, a strategy that continuously replaces existing data samples with newly generated pairs. This dynamic approach enhances adaptability by progressively exposing the model to an extensive dataset reflecting diverse optical conditions. Throughout training, the model processed approximately 650,000 unique data pairs across 130 iterations of incremental learning. Each iteration consisted of 50 epochs of training on mini-batches of 32 samples. We employed the Adam optimizer~\cite{Kingma2015}, which updates the model weights efficiently using adaptive moment estimation, starting with an initial learning rate of~$10^{-4}$. Training progress was monitored using validation data comprising 25\% of the generated samples.

In order to have a continuous dependence of the training loss on the point-like emitter positions in the ground truth, we refined the conventional mean squared error loss by applying a Gaussian filter to both the predicted and ground truth images. This step reduces sensitivity to slight reconstruction misalignments, improving the stability and overall training efficiency. Additionally, we incorporated an entropic regularization term into the loss function, controlled by the $\epsilon = 10^{-5}$ parameter, whose value was chosen experimentally. This regularization promotes sparsity of the output, encouraging the model to produce point-like emitters in the reconstructed image. The final loss averaged over an $N$-sampled batch is given by 
$\frac{1}{N} \sum_{i=1}^{N} \Vert I_i \ast G - \hat{I}_i  \ast G \Vert_{2}^{2} + \epsilon E(\hat{I}_i)$, 
where $I_i$ and $\hat{I}_i$ represent the target ground truth and predicted reconstruction, respectively, $G$ is the Gaussian filter, and $E(\hat{I}_i) = \hat{I}_i \log(\hat{I}_i)$ denotes the entropic regularization term.


\section{Quantum dot samples for deep-learning model testing}

To test our deep-learning model, we used three different quantum emitter samples: In(Ga)As quantum dots embedded in a planar cavity, prepared in sample~\textbf{1} with low 
($4.87\times10^{-3}\,\mu\text{m}^{-2}$) 
and in sample~\textbf{2} with high 
($2.05\times10^{-2}\,\mu\text{m}^{-2}$) 
quantum dot density, and a quantum emitter sample~\textbf{3} formed in a monolayer of tungsten diselenide (WSe$_2$), a type of transition metal dichalcogenide (TMD). 

We first verified the performance of our deep-learning model using the low-density In(Ga)As quantum dot sample, where the emitters were well isolated. We selected a site containing four well-separated quantum dots and acquired an image using a standard CMOS camera. We intentionally reduced the signal-to-noise ratio to test the capabilities of the model. In this sample, the quantum dots were embedded in a planar cavity structure composed of alternating AlAs/GaAs mirror pairs with 18~$\lambda$/4-thick bottom layers and 5 top layers, spectrally resonant with the quantum dot emission at 910~nm. The deep learning reconstruction results are presented in Sec.~\ref{Results}.

After confirming that our model accurately reconstructed images under conditions where no quantum dots overlapped, we proceeded to the high-density quantum dot sample, where partially overlapping quantum dots made Gaussian localization incapable. The second sample employed a planar cavity heterostructure similar to the low-density case but with a different mirror configuration. Namely, the cavity with $\lambda$ thickness, resonant with the quantum dot emission at 932~nm, comprised 24 bottom and 5 top layers of AlGaAs/GaAs mirrors. Both types of heterostructures were fabricated by molecular beam epitaxy, with details of the growth process provided in~\cite{Maier2014}.

Finally, we challenged the model using the WSe$_2$ TMD sample, which typically provides harsh imaging conditions due to its low signal-to-noise ratio and inhomogeneous background noise. The quantum dots in the WSe$_2$ monolayers were formed by strain~\cite{tripathi2018spontaneous, turunen2022quantum}. To accomplish this, a SiO$_2$ substrate was coated with a 200~nm-thick layer of gold. After coating, a protection layer of Al$_2$O$_3$ with a thickness of 5~nm was deposited. Several 100~nm deep square holes with varying side lengths have been etched in the Au/Al$_2$O$_3$ layer with a focused ion beam. Atomically thin flakes were mechanically exfoliated from a bulk WSe$_2$ crystal using PVC tape and transferred onto a structured substrate surface by the dry-gel stamp method~\cite{CastellanosGomez2014} at a temperature of 60$~^{\circ}$C. Nanoindentation of a flake transferred onto such structures causes local strain to arise in the monolayer and, in turn, creates quantum dot-like potential traps for excitons. As a result, strongly localized excitonic states are formed, manifesting themselves as quantum emitters, with emission wavelengths in the range of 720-760~nm.


\begin{figure}
	\includegraphics[width=0.8\columnwidth]{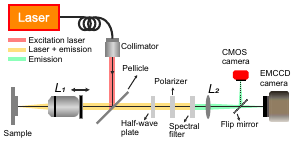}
	\centering
	\caption{
		Schematic illustration of the optical setup used for imaging quantum dots. A collimated excitation beam is directed onto the sample via a pellicle beamsplitter and focused by an aspheric lens ($L_1$). The same lens also collects the emitted photons, followed by polarization and spectral filters, which suppress residual excitation light. A second lens ($L_2$) then refocuses the filtered emission onto either a CMOS or an EMCCD camera for detection.}
	\label{fig:setup}
\end{figure}

\section{Experiment setup for quantum dot imaging}

In order to observe the emission from the quantum dots, the samples were placed into a closed-cycle cryostat maintained at 4.8~K. In(Ga)As quantum dots were excited using a pulsed Ti:sapphire laser at 810~nm, while the WSe$_2$ quantum dots were excited using a continuous-wave semiconductor laser at 638~nm. As shown in Fig.~\ref{fig:setup}, the collimated, linearly polarized laser beam was focused onto the sample via an aspherical lens $L_1$, which was also used to collect the emitted photons. The residual excitation laser beam was filtered out using a combination of a Glan-Thomson polarizer, a half-wave plate, and a long-pass filter, with a cut-off at 850~nm for the In(Ga)As quantum dot samples and 715~nm for the WSe$_2$ quantum dot samples.

The collected photons were focused by the lens $L_2$ and directed either to a CMOS camera (pixel size 9.25~$\mu$m~$\times$~9.25~$\mu$m) or an EMCCD (pixel size 20~$\mu$m~$\times$~20~$\mu$m), as shown in Fig.~\ref{fig:setup}. The EMCCD camera provides higher sensitivity and signal-to-noise ratio, while the choice between the two was made depending on the desired image quality. The focal lengths of both the aspheric lens $L_1$ and the lens $L_2$ preceding were chosen based on the desired magnification for each sample. 

For recording the images, the position of the aspheric lens $L_1$ was optimized to focus the emitted photons onto the image plane of the camera. All acquired images were taken in $50 \times 50$ pixels of the region of interest. To test the performance of our deep-learning model under various signal-to-noise ratios, we varied the excitation laser power and exposure time.


\section{Results} \label{Results}

We evaluated the performance of our deep-learning model using experimentally acquired images. To demonstrate the universality of the model, we analyzed results obtained under three distinct imaging conditions. Using a single low-resolution camera-captured intensity frame per sample, the model produces high-resolution reconstructions without requiring prior information on optical parameters or specialized training tied to an individual imaging system. All provided scale bars and absolute distances were obtained by comparing analytical calculations with numerical estimates of the Rayleigh resolution limit.

\begin{figure}
	\includegraphics[width=0.85\columnwidth]{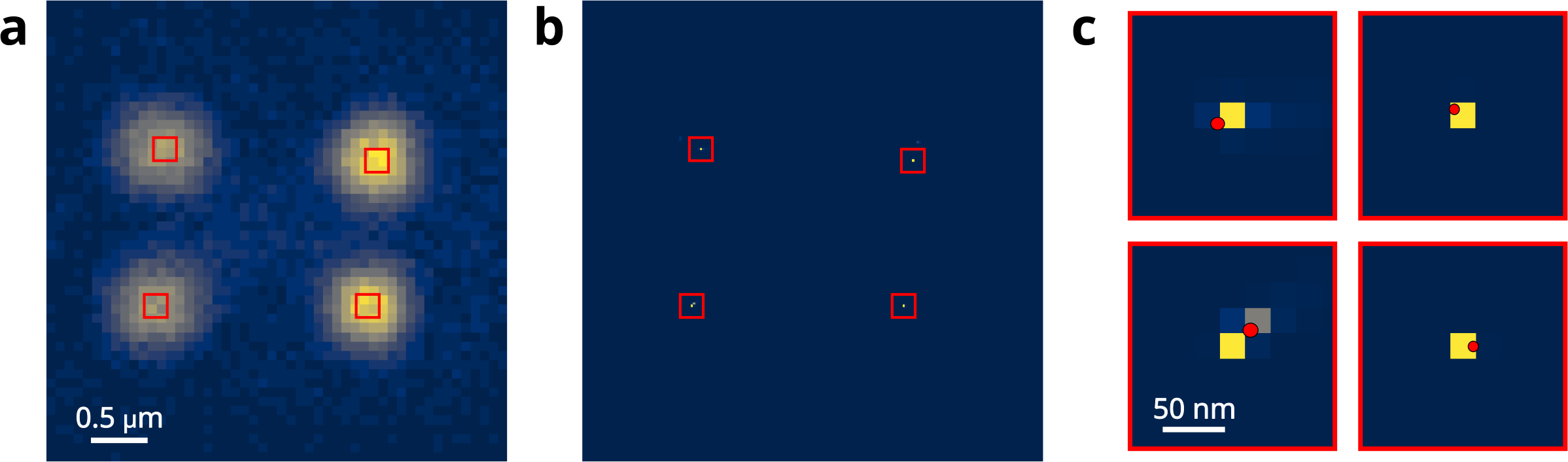}
	\centering
	\caption{
		Demonstration of deep-learning model performance on a sparse quantum dot image. \textbf{a} A low-resolution intensity image containing four well-separated quantum dots with a low signal-to-noise ratio. \textbf{b} The super-resolved reconstruction provided by the model with no prior information on the imaging system. The red rectangles indicate regions of interest that are magnified as insets in panel \textbf{c}. The zoomed-in view of the selected areas, with red circles denoting the positions and uncertainties estimated by a localization technique, i.e., the 2D Gaussian fitting. The discrepancy of 11~nm between the predictions of our deep learning model and the fitting-based localization is 72~times smaller than the Rayleigh optical resolution limit of 0.79~$\mu$m, demonstrating their excellent agreement.
	}
	\label{fig:5_1}
\end{figure}

First, we characterized the performance using a sparse sample containing four isolated quantum dots, as shown in Fig.~\ref{fig:5_1}. Panel \textbf{a} presents the low-resolution camera image acquired under low signal-to-noise ratio conditions. Despite visible astigmatism in the field of view, the model successfully reconstructed a highly accurate super-resolved image, see panel \textbf{b}. To better illustrate the improvement, panel \textbf{c} provides zoomed-in views of selected regions of interest. Because the quantum dots are clearly separable in the low-resolution image, we also applied a localization technique by fitting a 2D asymmetric Gaussian function to each emitter. Their estimated positions and associated uncertainties are marked as red circles in panel \textbf{c}. On average, the distance between the localization positions and the reconstruction centers of emitters obtained by the deep-learning model is approximately 11~nm, which is about 72~times below the 0.79~$\mu$m Rayleigh resolution limit of the imaging setup. These results highlight not only the ability of the universal model to significantly improve image resolution but also its high precision regarding the positions of the quantum dots.

\begin{figure}[ht!]
	\includegraphics[width=0.8\columnwidth]{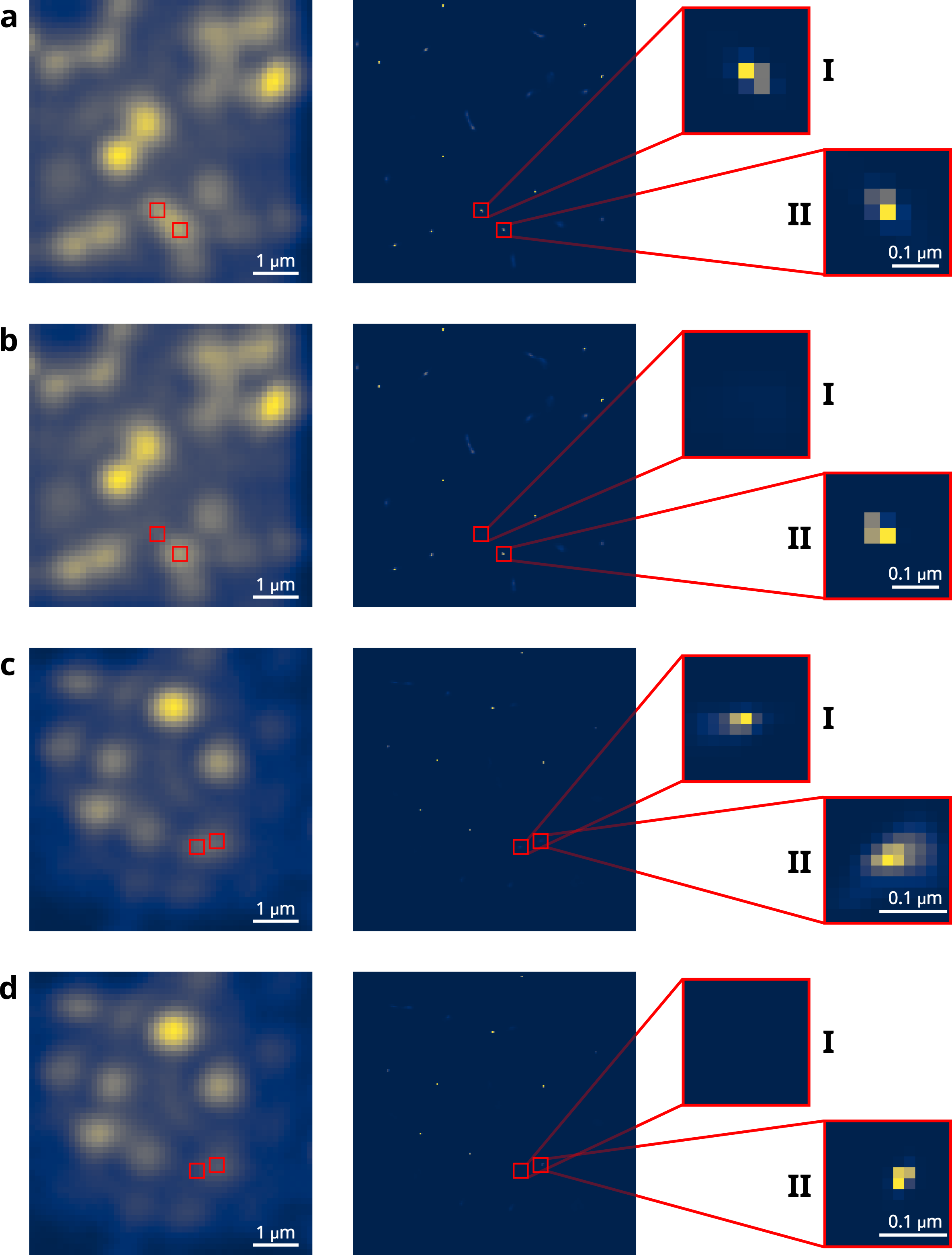}
	\centering
	\caption{
		Super-resolved reconstructions of two high-density images of overlapping quantum dots with the 0.81~$\mu$m Rayleigh resolution limit. Both panels \textbf{a} and \textbf{b} depict the same sample region but were captured at different time points, similarly to panels \textbf{c} and \textbf{d}. While the overall structural features remain consistent between the corresponding frames, a subtle change occurs in the lower area of panels \textbf{b} and \textbf{d}, where one quantum dot ceases to emit light. The reconstructions are virtually identical outside this region, which demonstrates the stability of the model. The areas highlighted with red squares indicate the change and are shown in detail in the zoomed-in insets \textbf{I} and \textbf{II}, where the disappearance of the emitter between frames is clearly resolved.
	}
	\label{fig:5_2}
\end{figure}

Since sparse sample reconstruction can be readily achieved with localization techniques, we next evaluated our model on a highly dense sample containing overlapping quantum dots. While localization-based approaches perform well for sparse samples, they fail in dense regions where multiple objects overlap. By contrast, our model processes the intensity image directly and provides high-resolution reconstructions even in such challenging conditions, see Fig.~\ref{fig:5_2}. In order to find out the ground truth (for a part of the sample) and evaluate the resolution enhancement, we utilized the blinking of a single quantum dot. It is important to stress that quantum dots elaborated for quantum technologies are stable sources, as their blinking would have a detrimental effect when used in quantum technology applications. Therefore, one cannot rely on their temporal fluctuations to improve image resolution. However, during our measurements, we observed two distinct emitters that featured apparent blinking. Panels \textbf{a} and \textbf{b} of Fig.~\ref{fig:5_2} show the same sample at two different time points, similarly to \textbf{c} and \textbf{d} depicting another region at other time points. Starting with the first two panels, these images reveal a structural change in the lower region of the field of view that suggests blinking of an emitting quantum dot. We reconstructed and compared these two frames, focusing on a pair of quantum dots marked in red in Fig.~\ref{fig:5_2} \textbf{a}. As shown in the insets, one quantum dot completely disappeared between frames (\textbf{I}), while the other remains virtually unchanged (\textbf{II}), along with the rest of the image. These reconstructed frames exhibit a normalized cross-correlation of~0.96 and a structural similarity index measure~\cite{ZhouWang2004} of~0.997 outside the blinking region, demonstrating the high stability and reproducibility of the reconstructions. Moreover, exploiting this temporal information, we localized the disappearing dot by analyzing the difference between frames. The localized position lies approximately 90~nm from the network-reconstructed center of the emitter, which is 9 times below the 0.81~$\mu$m Rayleigh limit. We also performed the corresponding analysis for the images shown in panels \textbf{c} and \textbf{d} and obtained a cross-correlation of~0.98 and a structural similarity index measure of~0.998. Similarly, the localized quantum dot position differs from the reconstructed center of mass by 78~nm, which is 10 times below the Rayleigh limit. It is also important to stress that the pairs of dots resolved in our reconstruction are overlapping (not resolved) in the original images. For example, the two dots under consideration in Fig.~\ref{fig:5_2} \textbf{c} are 1.6 times below the Rayleigh resolution limit. These results demonstrate the ability of our model to resolve individual quantum dots in dense environments and its sensitivity to subtle temporal dynamics.

\begin{figure}
	\includegraphics[width=0.7\columnwidth]{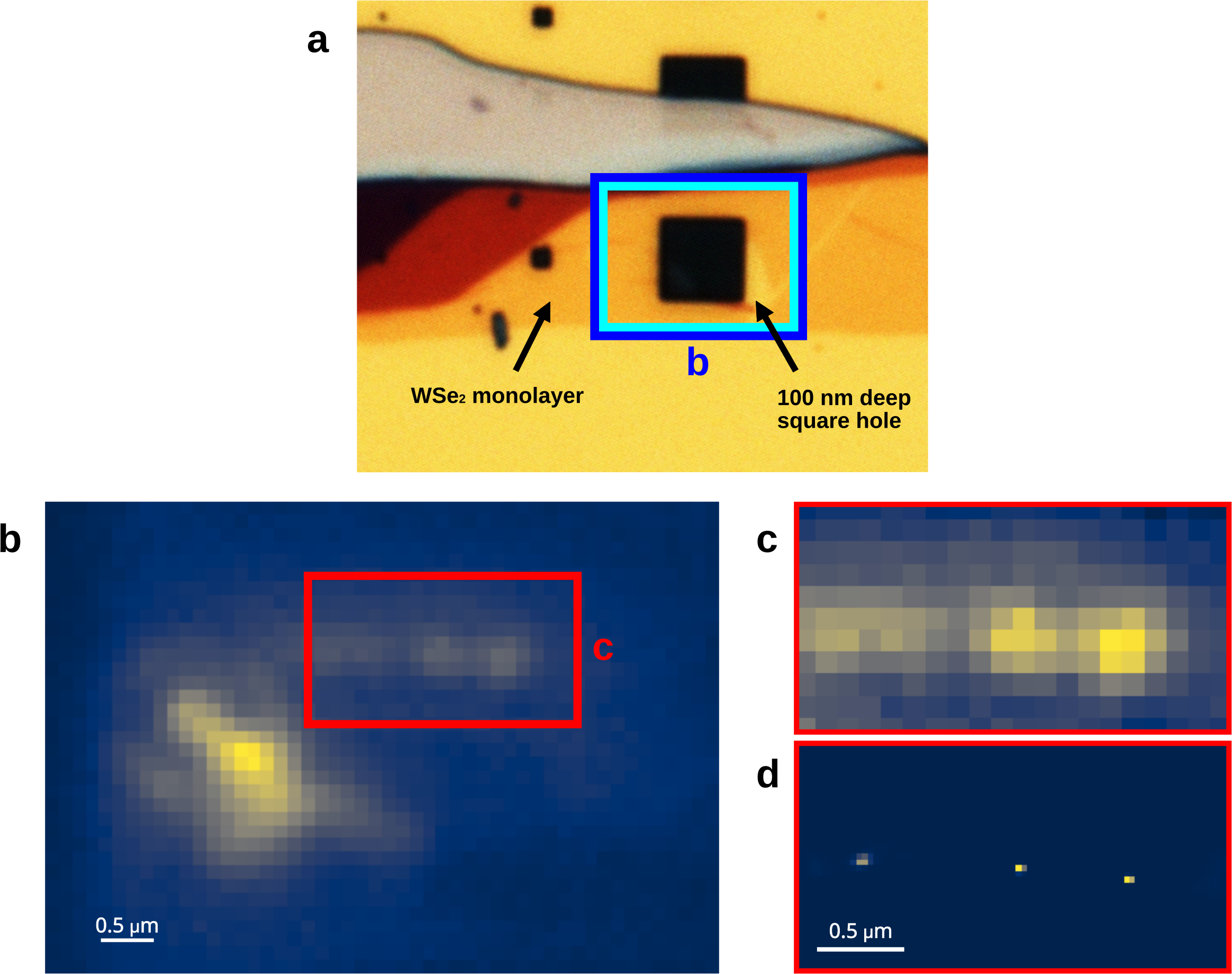}
	\centering
	\caption{
		\textbf{a} Wide-field image of a TMD monolayer quantum dot sample. The WSe$_2$ monolayer exfoliated from a nanoflake contains several etched, 100~nm deep square holes. The blue highlight indicates the region where the quantum dot image was taken. \textbf{b} Quantum dot emission image. Strongly localized excitonic states formed along the edge of an etched square hole behave as quantum dots. This low-resolution image poses a challenging case due to the combined effects of low signal-to-noise ratio and background inhomogeneity. \textbf{c, d} Zoomed-in view of the quantum dots and their super-resolution reconstruction. Despite the challenging conditions, the model successfully reconstructs a high-resolution image, revealing multiple quantum dots aligned along the edge. These reconstructed quantum dots form a straight line and are separated by approximately 0.85~µm (left to middle) and 0.59~µm (middle to right).
	}
	\label{fig:5_3}
\end{figure}

Lastly, to test the limits of our approach, we applied the model to a particularly challenging image of quantum dots aligned along the edge of a WSe$_2$ nanoflake. As shown in Fig.~\ref{fig:5_3}, this configuration presents a complex scenario combining a low signal-to-noise ratio with an inhomogeneous background arising from the nanoflake structural features. In the low-resolution camera image, the underlying pattern is barely recognizable. It is worth noting that such inhomogeneities were never encountered during training, as they were not part of the data simulation. Nevertheless, using our model, we successfully reconstructed a high-resolution image that clearly reveals three distinct quantum dots positioned along the nanoflake boundary. The reconstructed quantum dots are separated by approximately 0.85~$\mu$m and 0.59~$\mu$m, forming a straight line. A least-squares fitted line to their positions yields an average perpendicular deviation of only 9~nm, which is significantly below the Rayleigh resolution limit of 0.65~µm. These findings highlight the ability of our model to recover fine structural details under adverse imaging conditions, further underscoring its robustness and application potential across diverse experimental scenarios.

Overall, the universal model demonstrates exceptional super-resolution performance across all three configurations examined. It can accurately reconstruct high-resolution images and operate in highly challenging conditions, such as dense samples, low signal-to-noise ratio, and complex backgrounds. These results clearly demonstrate the versatility of the model, as it provides reconstruction from a single camera frame and requires no calibration or prior information about the imaging setup.


\section{Conclusion}

We introduced a calibration-free, single-shot computational super-resolution framework for imaging dense samples of quantum emitters. Trained entirely on synthetically generated images that span broad optical conditions, our convolutional model reconstructs super-resolved emitter maps directly from a single intensity frame, without requiring prior knowledge of the imaging system, spectral selectivity, or stochastic blinking. We validated the method across three distinct samples, sparse and dense In(Ga)As quantum dots in planar cavities and strain-induced emitters in WSe$_2$ monolayers, covering low signal-to-noise ratios, overlapping PSFs, astigmatism, and inhomogeneous backgrounds. Despite being acquired under different imaging conditions, the images from all three samples are super-resolved using the same deep-learning model without retraining or fine-tuning, demonstrating its universality. In all cases, the model produced precise localizations well below the corresponding Rayleigh limits and recovered spatial configurations that were inaccessible to conventional single-frame fitting. Because the developed framework is universal and fast, it is immediately useful as a metrology and feedback tool for nanophotonics and quantum-device fabrication, where accurate inter-emitter spacing and alignment determine device performance. We believe our approach will enable universal, single-shot super-resolution across a wider class of solid-state emitters, supporting scalable quantum-photonic technologies and related nanoscale science.


\section*{Supplementary Information}
The following supplementary information is available in a standalone document: a detailed analysis of generalization ability beyond the imaging conditions encountered during training.

\section*{Funding}
Ministry of Education, Youth, and Sports of the Czech Republic (project OP JAC CZ.02.01.01/00/23\_021/0008790);
Wallenberg Centre for Quantum Technology (WACQT);
Swedish Research Council grant 2021-04494;
Czech Science Foundation (project 21-18545S);
Palacký University Olomouc (projects IGA-PrF-2025-010, IGA-PrF-2026-005);
German Federal Ministry of Research, Technology and Space (projects COMPHORT and TubLan);
QuantERA (project COMPHORT).

\section*{Acknowledgment}
J.L. and C.St. were supported by the Knut \& Alice Wallenberg Foundation (through the Wallenberg Centre for Quantum Technology (WACQT)). A.P. would like to acknowledge the Swedish Research Council (grant 2021-04494).
D.V. and M.J. acknowledge the use of cluster computing resources provided by the Department of Optics, Palack\'{y} University Olomouc. We thank Jan Provazn\'{i}k for maintaining the cluster and providing support.\\ 
This project was funded within the QuantERA programme that has received funding from the European Union's Horizon 2020 research and innovation programme under Grant Agreement No. 101017733 (Project COMPHORT), and with funding by the German Federal Ministry of Research, Technology and Space (BMFTR) within the projects COMPHORT and TubLan.

\section*{Disclosures}
The authors declare no conflicts of interest.

\section*{Data Availability}
The code and data that support the findings of this study are publicly available on GitHub: \\\url{https://github.com/VasinkaD/Quantum-Dots-Super-Resolution}

\section*{Author Contributions}
D.V. developed deep-learning models and performed numerical simulations and data processing. J.L. and C.St. developed the optical experiment and performed data acquisition. A.P. supervised the experimental part of the project. M.J. conceived the idea of universal image super-resolution and supervised the theoretical part of the project. S.H. and C.Sc. provided samples \textbf{1} and \textbf{2}. V.M., I.S., S.S., F.E., S.A.T., and C.Sc. fabricated sample number \textbf{3}. D.V., M.J., A.P., J.L., and V.M. wrote the manuscript, and all authors were involved in revising the manuscript. 


\end{document}